\documentclass[aps,pre,twocolumn, groupedaddress]{revtex4-2}

\usepackage{graphicx}
\usepackage{xcolor, amsmath, comment}

\begin{document}
\title{Collective Turns in Spinless Flocks}
\author{Joao Lizárraga}
\author{Marcus A. M. de Aguiar} \email{aguiar@ifi.unicamp.br}
\affiliation{
	Instituto de Física Gleb Wataghin, Universidade Estadual de Campinas, Unicamp 13083-970, Campinas, São Paulo, Brazil
}
\date{\today}

\begin{abstract}
Using a minimal aggregation-based model, we address the efficient information transfer observed in natural flocks during collective turns. In our system, the turning signal emitted by a single individual propagates throughout the collective primarily at a constant speed. To characterize the source of this capability, we analyze the system in the continuum limit. Thus, we demonstrate that wave-propagating features arise solely from the non-reciprocal nature of interactions. We explore spectral anomalies that emerge due to this mechanism and provide an explanation through a perturbative analysis. Remarkably, within this scenario, the description of velocity fluctuations is found to follow a structure analogous to a Born approximation. Our model can be extended to study other physical contexts exhibiting polar order, and our arguments can be tested in natural systems.
\end{abstract}
\maketitle

\section{Introduction}
Birds gather and fly together, forming a flock. The early observation of this phenomenon led to the study of such collectives using frameworks based on a velocity alignment mechanism. Indeed, for a long time, the Vicsek model~\cite{vicsek1995novel, chate2008collective, peruani2008mean} was considered a pillar in the study of related problems~\cite{barberis2016large, miguel2018effects, codina2022small, gao2025swarming}. This status, however, faced challenges as subsequent field observations were carried out. Particularly, studies on starling flocks revealed hallmarks of their collective behavior that were not contemplated by Vicsek's formulation: strong alignment and cohesion are maintained during both straight flight and turns~\cite{ballerini2008interaction, cavagna2010scale}. Moreover, despite the absence of leaders to reach collective coherence, the flocks' turning events were found to be triggered by single individuals~\cite{attanasi2014information}.

After recognizing the limitations of the Vicsek model in reproducing the phenomena exhibited by starlings, Cavagna \textit{et al.} introduced the Inertial Spin Model (ISM)~\cite{cavagna2015flocking, cavagna2024discrete}. At its core, the ISM stems from the incompleteness of the Vicsek model. It does rely on the velocity-alignment framework; yet, rather than directly promoting the alignment of individual velocities, it does so through internal rotation generators: the \textit{spins}. As argued by Cavagna \textit{et al.}, the Vicsek model characterizes an overdamped system: if waves emerge, the system's nature causes them to dissipate rather than propagate. In contrast, the ISM allows the system to operate in one of two regimes: overdamped or underdamped. Therefore, since real starling flocks were shown to propagate waves during collective turns~\cite{attanasi2014information}, they were hypothesized to match the ISM underdamped scenario.

Although the ISM effectively reproduces the turning phenomenon of starling flocks, recent experiments have given rise to an apparent \textit{paradox}~\cite{cavagna2025spin}. Under the ISM formulation, if the flock were underdamped, two peaks, representing the spin-waves, should emerge in the frequency domain of velocity fluctuations. However, in real starling flocks, these peaks are not observed. Thus, to reconcile the experimental evidence with the theoretical framework, Cavagna \textit{et al.} proposed that starlings may operate adaptively depending on the amplitude of fluctuations. That is, for small fluctuations, the system is overdamped, and for larger ones (turning signals), the system becomes underdamped.

In this paper, we address the propagation phenomenon experienced by turning flocks without explicitly relying on the \textit{standard} Vicsek-like alignment framework. Instead, we extend a model of aggregation introduced in~\cite{lizarraga2025swarming}, which characterizes a collective of individuals who align their velocities as a by-product of achieving cohesion. The dynamics of this extended system are discussed in Sec.~\ref{sec:model}, where we also illustrate its behavior during collective turns. Subsequently, in Sec.~\ref{sec:info}, we explore how the turning information propagates across the collective. This analysis continues in Sec.~\ref{sec:cont}, where we address the system's nature from a continuum perspective; related numerical evaluations are presented in Sec.~\ref{sec:corr}. Finally, in Sec.~\ref{sec:dis}, we summarize our findings and discuss their broader implications.

\section{The model}\label{sec:model}
In the aggregation model presented in~\cite{lizarraga2025swarming}, each individual's self-propelling velocity is constant. Here, in contrast, we consider that individuals are affected by their intention to match neighboring velocities. That is, self-propulsion is \textit{adaptive}. Thus, a system of $N$ individuals is governed by 
\begin{subequations}
\label{eq:iner}
\begin{align}
    \dot{\vec{r}}_i &= \vec{f}_i(t) + \alpha\left(\langle\vec{r}\rangle - \vec{r}_i\right),\label{eq:agg}\\
    \dot{\vec{f}}_i &= \vec{\nu}_i(t) + \gamma\sum_{j =1}^N A_{ij}(\dot{\vec{r}}_j - \dot{\vec{r}}_i),\label{eq:ali} 
\end{align}
\end{subequations}
where $\vec{r}_i$ and $\vec{f}_i(t)$ respectively represent the $i$-th individual's position and self-propelling velocity. The random variable, $\vec{\nu}_i(t)$, is characterized by $\langle\vec{\nu}_i(t)\rangle = 0$ and $\langle\vec{\nu}_i(t)\vec{\nu}_i(t')\rangle = 2D_{\nu}\delta(t-t')$, and $\alpha$ and $\gamma$ are constant coupling strengths. Local interactions are ruled by the adjacency matrix $A$: the self-propulsion of the $i$-th individual evolves affected by the state of its $\eta$ closest neighbors in space. This restriction thus implies non-reciprocity, a feature that manifests itself through the non-symmetry of $A$.

\subsection{Microscopic dynamics}
Formulating the model as in Eqs.~\eqref{eq:iner} enables us to recognize the dependence of the system's dynamics on two core components: aggregation [Eq.~\eqref{eq:agg}] and alignment [Eq.~\eqref{eq:ali}]. The expected collective behavior can thus be understood by analyzing the individual and combined effects of both mechanisms.

From Eq.~\eqref{eq:agg}, it is trivial to notice that the aggregation dynamics are stable if $\alpha>0$. The equilibrium is characterized by the absence of velocity fluctuations ($\vec{\varphi}_i = \dot{\vec{r}}_i - \langle\dot{\vec{r}}\rangle = 0$), thus describing a state of global alignment. Notably, as a consequence of reaching this condition, the relative positions adhere to the relationship 
\begin{equation}
    \vec{r}_i - \langle\vec{r}\rangle = \frac{1}{\alpha}(\vec{f}_i - \langle\vec{f}\rangle),
    \label{eq:posconv}
\end{equation}
which characterizes the system's cohesion. 

The aggregation equilibrium is easily achieved when individuals are defined by constant self-propelling velocities ($\vec{f}_i$). In this scenario, the aggregation coupling strength ($\alpha$) determines the relaxation time characteristic of the transient in which $\vec{\varphi}_i$ reaches zero. When $\vec{f}_i$ is time-varying, in contrast, rather than driving the asymptotic convergence of $\vec{\varphi}_i$ to zero, the aggregation mechanism promotes a tracking behavior. Velocity fluctuations will continuously approach zero over time, but will never strictly reach $\vec{\varphi}_i=0$. In spite of this, as long as $\alpha$ is positive and $\vec{f}_i(t)$ is bounded, the system remains cohesive.

Setting $\alpha = 0$ prevents us from monitoring the collective's spatial state; a relation such as Eq.~\eqref{eq:posconv} cannot be derived. However, the alignment mechanism [Eq.~\eqref{eq:ali}] alone does promote coherence ($\dot{\vec{r}}_i = \dot{\vec{r}}_j$). Consider, for instance, all-to-all interactions ($A_{ij} = 1$) and no aggregation ($\alpha = 0$). Thus, Eqs.~\eqref{eq:iner} lead to $\dot{\vec{f}}_i = \gamma(\langle\vec{f}\rangle - \vec{f}_i)$, which drives the convergence of $\vec{f}_i$ towards $\langle\vec{f}\rangle$ (for $\gamma >0$). In turn, since $\dot{\vec{r}}_i = \vec{f}_i(t)$, velocities align. A similar state is achieved when alignment interactions are topological, with fixed $\eta$, as formally considered in our model. In this scenario, however, a positive value of $\gamma$ drives the system's coherence at a local scale. Consequently, although both mechanisms aim for velocity coherence, setting $\alpha$ and $\gamma$ to be positive generates an interplay between local and global effects. 

Within the non-aggregating scenario, individuals may successfully reach global alignment and temporarily maintain a fixed spatial structure. However, moving in parallel is not equivalent to moving together. For $\alpha = 0$, Eqs.~\eqref{eq:iner} describe the evolution of velocities as an Ornstein–Uhlenbeck process on a network. After an initial transient, velocity differences ($\dot{\vec{r}}_j - \dot{\vec{r}}_i$) fluctuate around zero, yet, relative positions diffuse such that $\mathrm{Var}(\vec{r}_j - \vec{r}_i)\sim t$. Consequently, without the aggregation mechanism, the system is susceptible to evaporation. Introducing an explicit dependence on relative positions ($\alpha>0$) generates a \textit{restoring force} that ensures cohesion within fixed spatial boundaries.

Analyzing the role of the alignment mechanism requires particular care, as velocity coherence is promoted indirectly. Although the local dynamics depend on individual velocities via ($\dot{\vec{r}}_j - \dot{\vec{r}}_i$), they characterize the evolution of $\vec{f}_i$, not $\dot{\vec{r}}_i$. This intermediate stage ultimately provides the system with \textit{memory}: self-propulsion depends on a history of positions. Remarkably, the existence of $\dot{\vec{f}}_i$ implies that the system is governed by second-order dynamics. Nevertheless, given that accelerations ($\ddot{\vec{r}}_i$) are independent of absolute positions ($\vec{r}_i$), the dynamics governing the velocity fluctuations are essentially non-inertial.

\subsection{Collective turns}
As described, alignment and aggregation mechanisms drive the system towards a state of orientational coherence. Therefore, while affected solely by noise, the collective reaches a quasi-steady state of straight flight. This behavior, however, is challenged when the system experiences a sustained perturbation, which further reveals the interplay between local and global effects. 

To illustrate the system's behavior, we performed numerical simulations using Eqs.~\eqref{eq:iner}. In each execution, individuals first achieved a state of straight flight. Then, a turning event was triggered by varying the self-propelling velocity of a single individual (i.e., its intention). Specifically, we extracted the respective $\vec{f}_{initiator}$ at a specific time step and enforced its gradual rotation around the $z$-axis. This process was set to last for a fixed number of simulation steps. Remarkably, during the turn, the initiator's velocity remained regulated by the aggregation dynamics:
\begin{equation*}
\dot{\vec{r}}_{initiator} = \vec{f}_{initiator} + \alpha(\langle\vec{r}\rangle - \vec{r}_{initiator}).    
\end{equation*}
Computational details, including specific parameters, are provided in Appendix~\ref{app:num}.

\begin{figure*}
    \centering
    \includegraphics[width=\linewidth]{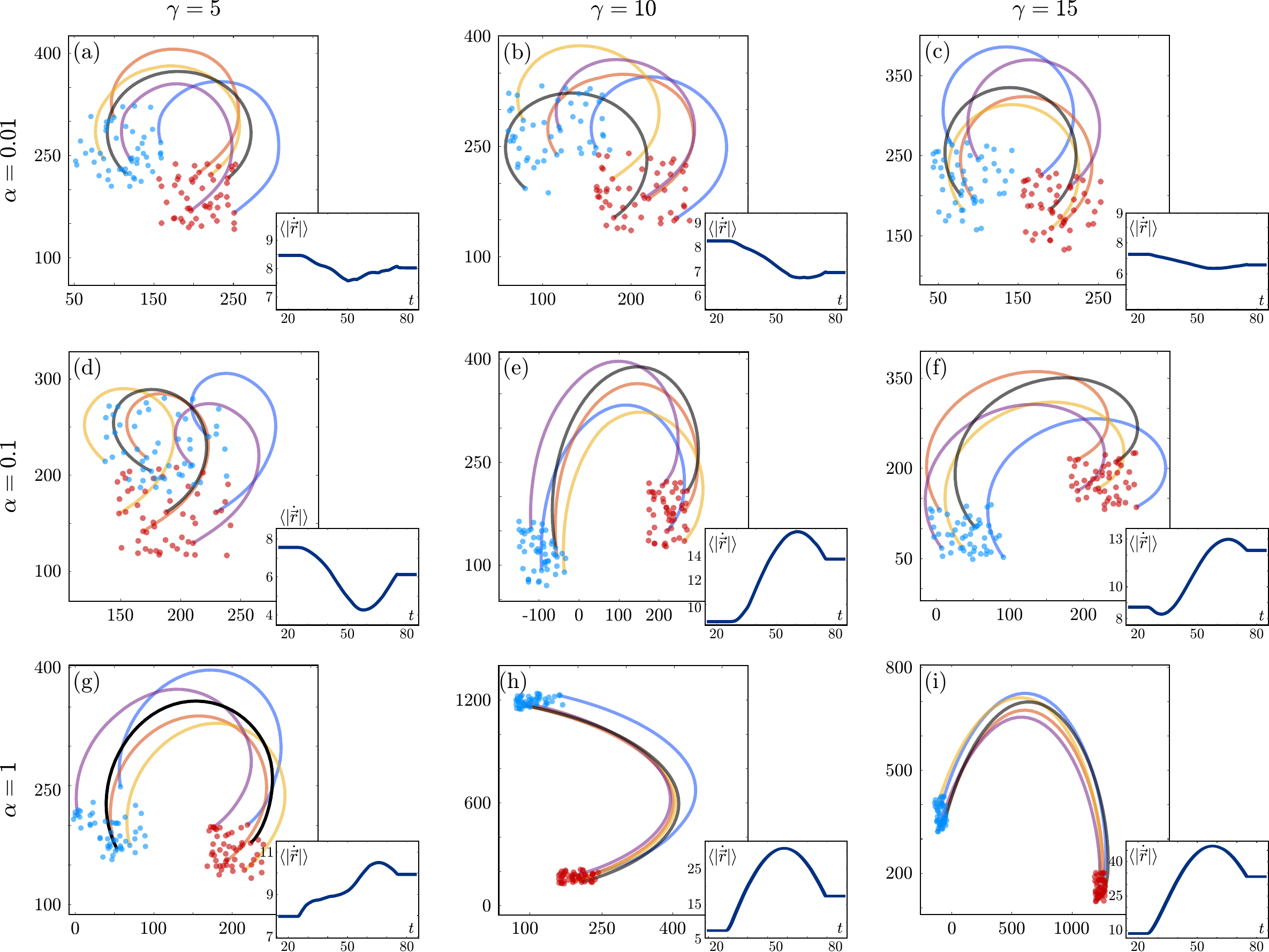}
    \caption{Spatial states reached in the $xy$-plane by groups of $N=50$ individuals during collective turns. Red dots represent the system's spatial distribution at the beginning of the turn; blue dots, at the end. In all cases, the initiator's self-propelling velocity ($\vec{f}_i$) was set to rotate $3\pi/2$ around the $z$-axis along $5\times10^4$ time-steps. Colored lines represent trajectories followed by five individuals (chosen at random), with the black one corresponding to the initiator's. The respective insets display the average speed calculated throughout each turning event. Coupling strengths are indicated on the left and top margins; for better visualization, $D_\nu$ was set at 0.}
    \label{fig:fig01}
\end{figure*}

As pictured in Figs.~\ref{fig:fig01}(a)-\ref{fig:fig01}(c), when $\alpha$ is small, the system responds predominantly to local dynamics. Once the initiator starts turning, its nearest neighbors seek for alignment. Then, non-reciprocal interactions associated with these neighbors perturb individuals beyond the initiator's locality. This process creates a chain of reactions which culminates on the alignment of individuals on a macroscopic level. Notably, even under the model's deterministic formulation ($D_\nu = 0$), velocity fluctuations emerge given that individuals experience competing alignment influences from distinct local clusters. The propensity for the development of these fluctuations is governed by $\gamma$, with larger values driving more rapid alignment, and vice versa.

A larger value of $\alpha$ leads to a contrasting behavior, as displayed in Figs.~\ref{fig:fig01}(d)-\ref{fig:fig01}(i). Rather than being dominated by local dynamics, the system becomes increasingly governed by its global state. In this scenario, the initiator's neighbors seek local coherence but are strongly influenced by the preferential orientation of the rest of the population. Consequently, if a collective turn is performed, individual trajectories are found to be nearly parallel. Note that this behavior differs from that shown in Figs.~\ref{fig:fig01}(a)-(c). In these, the relaxation of global interactions allows individuals to follow approximately equal-radii trajectories, aligning closely with observations of starling flocks~\cite{ballerini2008empirical, papadopoulou2023diffusion}. Finally, Figs.~\ref{fig:fig01}(d)-(f) depict states in an intermediate scenario: global interactions exert an influence, yet local effects are also significant.  

The insets in Fig.~\ref{fig:fig01} display how the collective's average speed evolves in each event. Consistent with observations in natural flocks~\cite{cavagna2022marginal}, equal-radii collective turns are characterized by minor speed variations. Remarkably, the relation between this behavior and the magnitude of $\alpha$ can be explained through the aggregation dynamics. Note, from Eq.~\eqref{eq:agg}, that individual speeds are characterized by
\begin{equation*}
    |\dot{\vec{r}}_i| = |\vec{f}_i + \alpha(\langle\vec{r}\rangle - \vec{r}_i)|.
\end{equation*}
That is, even under consistently high polarization, individual speeds are influenced by variations in relative positions scaled by $\alpha$. During collective turns, therefore, when individuals are not constrained to maintain a fixed $(\vec{r}_i - \langle\vec{r}\rangle)$, the effects of $\alpha$ become particularly noticeable. When $\alpha = 0.01$, speed variations are small, implying also that individuals closely follow the initiator's expected path. Their trajectories thus remain almost exclusively along the $xy$-plane. Large speed variations, in contrast, such as those emerging for $\alpha = 1$, characterize individuals that experience considerable accelerations while turning. As a consequence, they follow longer trajectories and even deflect along the $z$-axis.

\section{Information propagation}\label{sec:info}
The system's global coherence is maintained during collective turns, even when local effects are set to be more significant than global ones [Figs.~\ref{fig:fig01}(a)-\ref{fig:fig01}(c)]. To further characterize this behavior, we now focus on how the initiator triggers such a macroscopic state. 

\begin{figure}
    \centering
    \includegraphics[width=\linewidth]{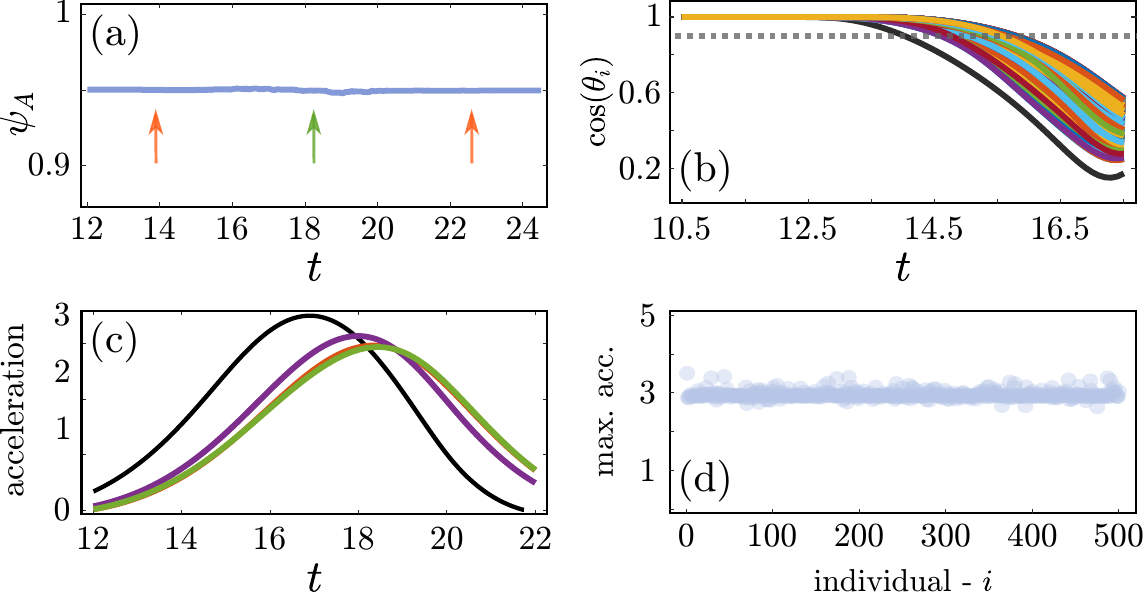}
    \caption{Signatures of information propagation in a system with $(N, \alpha, \gamma, D_\nu) = (500, 0.01, 50, 0.5)$. In (a), the asymmetry index, ranging from $0$ (perfect asymmetry) to $1$ (perfect symmetry), is calculated via $\psi_A = \|0.5(A+A^T)\|_F/ \| A\|_F$. The flock follows a sequence of aggregation (not shown), straight flight (orange arrow), turning (green arrow), and return to straight flight. In (b), the turning state of each individual is displayed as the evolution of the respective orientational index over time; the black line corresponds to the initiator's, and the dashed line represents the threshold $\cos(\theta_i) = 0.9$. The radial accelerations shown in (c) correspond to those of the initiator (black) and three random individuals during the turn. Maximum accelerations reached by all individuals are shown in (d).}
    \label{fig:fig02}
\end{figure}

The collective we evaluate follows the stages described in Fig.~\ref{fig:fig02}(a). As expected from Eq.~\eqref{eq:iner}, given the topological origin of non-reciprocity, the interaction matrix remains non-symmetric regardless of the system's state. During straight flight, all individuals share the same passive role; however, while turning collectively, their states diverge [Figs.~\ref{fig:fig02}(b)-\ref{fig:fig02}(d)]. Once one individual initiates the turning event, information about its change of state is received by local neighbors. Then, these react with a similar intensity, promoting in turn the response of their own neighbors. The propagation of the turning signal, almost lossless, is evidenced by the similarity between the individuals' radial accelerations [Figs.~\ref{fig:fig02}(c) and \ref{fig:fig02}(d)], which characterize their turning response~\cite{attanasi2014information}. 

\subsection{Front speed}
The initiator's turning signal reaches the other individuals in a hierarchical manner. Its speed of propagation can thus be estimated by calculating the time it takes for each individual to turn after the initiator~\cite{attanasi2014information,cavagna2015flocking}. 

For any particle, its current and initial (before the turn) velocities relate via: $\dot{\vec{r}}_i(t)\cdot\dot{\vec{r}}_i(t_0) = |\dot{\vec{r}}_i(t)||\dot{\vec{r}}_i(t_0)|\cos\theta_i(t)$. Then, given our focus on highly polarized collectives, we assume negligible speed variations. Individual turning states are thus characterized by $\cos\theta_i(t)\propto\dot{\vec{r}}_i(t)\cdot\dot{\vec{r}}_i(t_0)$ [Fig.~\ref{fig:fig02}(b)]. In our numerical analyses, a particle is considered to have successfully followed the initiator's signal once it reaches the threshold $\cos\theta_i(t = \bar{t}_i) = 0.9$. On this basis, the turning rank ($\tilde{r}_i$) is determined in consistency with the respective delays $\bar{t}_i$. When presenting our results, we rather use absolute delays ($t_i = \bar{t}_i - \bar{t}_{initiator}$) to facilitate comparison. 

As the turning signal originates from a localized source, the propagation of information is conceptualized as a sphere that grows over time. Particularly, given the system's dimensionality, the sphere's radius can be defined as $[\tilde{r}(t)/\tilde{\rho}]^{1/3}$, where $\tilde{r}(t)$ represents the number of individuals that have turned at instant $t$ and $\tilde{\rho}$ is the collective's spatial density. For simplicity, in all displayed figures we used $\bar{r}(t) = \tilde{r}(t)^{1/3}$, which does not induce or suppress regimes important for our analysis.

\begin{figure}
    \centering
    \includegraphics[width=\linewidth]{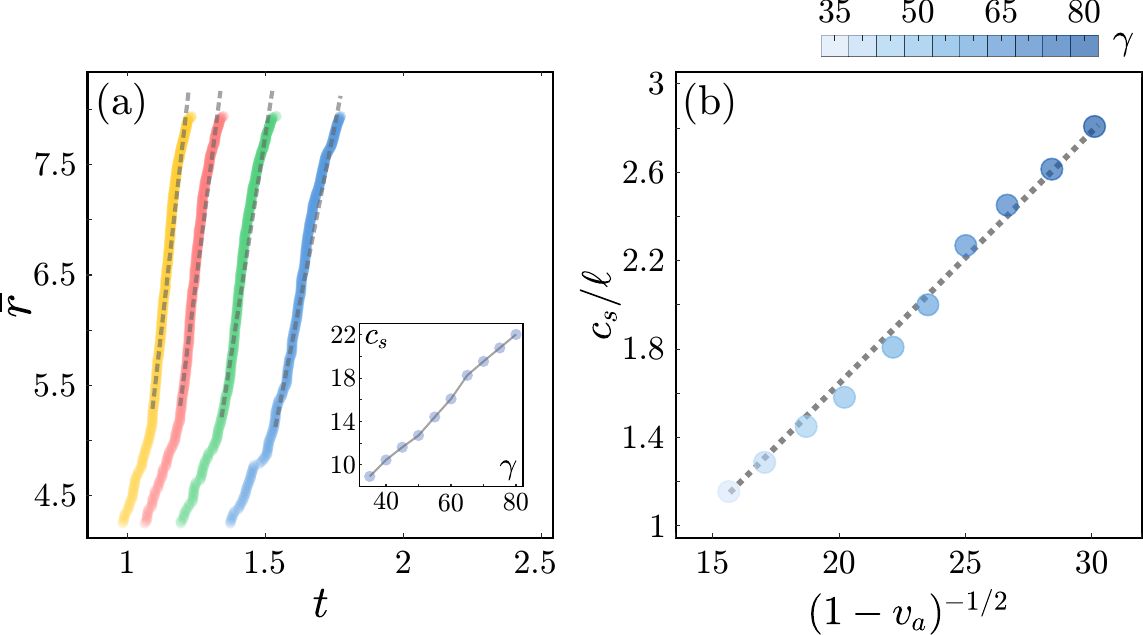}
    \caption{Propagation speeds estimated for systems with $(N, \alpha, D_\nu) = (500, 0.01, 0.5)$. In (a), the front speed ($c_s$) is estimated for $\gamma = 50$ (blue), $\gamma = 60$ (green), $\gamma = 70$ (red), and $\gamma = 80$ (yellow). The state of the collective before the turn was identical for all the cases. The dashed lines fit linear regimes; segments associated with the first turning individuals were cropped for better visualization. The inset shows the relation between $c_s$ and $\gamma$, including additional cases. In (b), the propagation speed is rescaled by the mean interparticle distance ($\ell$) and is shown to scale approximately linearly with the degree of polarization; the latter was calculated as $v_a = N^{-1}\left|\sum_i\left(\dot{\vec{r}}_i/|\dot{\vec{r}}_i|\right)\right|$. Each blue marker corresponds to data obtained for a different value of $\gamma$. }
    \label{fig:fig03}
\end{figure}

Addressing our system using the method described above yields the results shown in
Fig.~\ref{fig:fig03}. Successful collective turns give rise to regimes in which the distance traveled by the propagating signal ($\bar{r}$) scales linearly with time [Fig.~\ref{fig:fig03}(a)]. That is, the turning signal propagates at constant speed ($c_s$). As previously discussed regarding Eq.~\eqref{eq:ali}, larger values of $\gamma$ promote faster alignment. Therefore, since $c_s$ is estimated based on the state of individual orientations, it is expected to increase. The relation between $c_s$ and $\gamma$ is displayed in the inset of Fig.~\ref{fig:fig03} where, remarkably, the dependence is found to be approximately linear. Finally, in Fig.~\ref{fig:fig03}(b), we show that higher propagation speeds correlate with greater orientational coherence, which is a defining feature of natural flocks~\cite{attanasi2014information}. In Appendix~\ref{app:pp}, we elaborate an analytical argument on this matter. 
 
The procedure we employed to estimate propagation speeds in Fig.~\ref{fig:fig03} is essentially the same as that used in the study of starling flocks and the ISM~\cite{attanasi2014information, cavagna2015flocking}. Deviations from the linear regime, also observed in these works, are attributed to the collective's spatial boundaries (we show additional cases in Appendix~\ref{app:num}). Note, moreover, that if turning events were triggered at different times, and $\bar{r}$-$t$ relations were plotted for no absolute times (i.e., starting at $t\neq0$), over the entire time domain, most deviations from the linear regime would be imperceptible.

\section{The continuum limit}\label{sec:cont}
The ballistic-like propagation evidenced in Fig.~\ref{fig:fig03}(a) is consistent with that of real flocks. It is, indeed, one of the main reasons for treating them as wave-propagating (underdamped) rather than purely diffusive (overdamped) systems~\cite{attanasi2014information}. To explore the nature of our system in this context, we rely on its description in the continuum limit. Specifically, we focus on the behavior of velocity fluctuations ($\vec{\varphi}_i$), which are expected to propagate across the flock (acting as a medium).

We start by rearranging Eqs.~\eqref{eq:iner} into the fluctuation dynamics
\begin{equation}
    \dot{\vec{\varphi}}_i + \alpha\vec{\varphi}_i +\gamma\sum_{j = 1}^N L_{ij}\vec{\varphi}_j + \vec{\xi}_i= \vec{\nu}_i,
    \label{eq:fluc}
\end{equation}
where we neglect $\langle\vec{\nu}\rangle$, $L$ is the in-degree Laplacian associated with $A$, and $\vec{\xi}_i$ represents a stimulus that affects all the nodes uniformly. This term arises from the non-reciprocity of interactions, and can be written in terms of the nodes' in- and out-degrees as $\vec{\xi}_i = \gamma N^{-1}\sum_j\vec{\varphi}_j(d_j^{out} - d_j^{in})$. 

By extending Eq.~\eqref{eq:fluc} to the continuum limit, we obtain
\begin{equation}
    \dot{\vec{\mu}} + \alpha\vec{\mu} + \gamma\mathcal{L}\vec{\mu} + \vec{\xi} = \vec{\nu},
    \label{eq:fdyn}
\end{equation}
where $\vec{\mu}(\vec{r}, t)$ is the field of velocity fluctuations and $\mathcal{L}(\vec{r})$, the Laplacian operator, is characterized by (see Appendix~\ref{app:cont} for the full derivation):
\begin{equation}
 \mathcal{L} = \vec{C}_1\nabla + \nabla\cdot(\mathbf{C}_2\nabla).  
 \label{eq:lap}
\end{equation}
Here, the advection component depends on
\begin{equation*}
    \vec{C}_1(\vec{r}) = -\int d\vec{r}\,'\rho(\vec{r}\,')\mathcal{A}(\vec{r}, \vec{r}\,')(\vec{r}\,' - \vec{r}),
\end{equation*}
where $\mathcal{A}(\vec{r}, \vec{r}\,')$ characterizes a \textit{kernel} of interactions, and $\rho(\vec{r}\,')$ represents the local density at position $\vec{r}\,'$. When interactions are fully reciprocal, $\vec{C}_1$ vanishes. The diffusion component, on the other hand, depends on 
\begin{equation*}
    \mathbf{C}_2(\vec{r}) = -\int d\vec{r}\,'\rho(\vec{r}\,')\mathcal{A}(\vec{r}, \vec{r}\,')(\vec{r}\,' - \vec{r})\otimes (\vec{r}\,' - \vec{r}),
\end{equation*}
which, by construction, is negative semi-definite.

Note that the interaction network (and therefore $\mathcal{L}$) is expected to evolve during a collective turn. This is indeed supported by the small variations of the asymmetry index ($\psi_A$) observed in Fig.~\ref{fig:fig02}(a). However, in the remainder of this paper, we assume $\mathcal{L}$ to be time-invariant. This consideration does not undermine our subsequent results, but rather allows us to focus on the system's spatial dependence. Further simulations (not shown) verified that the effects discussed below are also present when the interaction network is constrained to be static.

\subsection{Spatial homogeneity}

Consider that $\vec{C}_1$ and $\mathbf{C}_2$ [Eq.~\eqref{eq:lap}] are constant across the collective. Then, for convenience, we rewrite Eq.~\eqref{eq:fdyn} as
\begin{equation}
    \dot{\vec{\mu}}_0 + \alpha\vec{\mu}_0 + \gamma\mathcal{L}_0\vec{\mu}_0 + \vec{\xi} = \vec{\nu},
    \label{eq:fdyn0}
\end{equation}
where we simply added a subscript to label the condition of spatial homogeneity. By performing the Fourier transform, we thus obtain $\hat{G}_0^{-1} \hat{\mu}_0= \hat{\nu}$, with  $\hat{.}$ denoting that the variable is in the frequency domain, and 
\begin{equation}
\hat{G}_0^{-1} = \alpha - \gamma\vec{k}\cdot\mathbf{C}_2\vec{k}+ i(\gamma\vec{C}_1\cdot \vec{k} - \omega).
    \label{eq:prop}
\end{equation}
Notice that transforming $\vec{\xi}$ yields a term proportional to $\delta(\vec{k})$, which vanishes as we restrict our study to cases where $\vec{k}\neq \vec{0}$. 

The deterministic scenario ($\hat{G}_0^{-1} \hat{\mu}_0= 0$) allows us to solve for the temporal frequency, leading to
\begin{equation}
    \omega(\vec{k}) = \gamma\vec{C}_1\cdot\vec{k} + i\left(\gamma\vec{k}\cdot\mathbf{C}_2\vec{k} - \alpha\right),
    \label{eq:disres}
\end{equation}
which characterizes the system's dispersion relation. Since $\mathrm{Re}[\omega(\vec{k})]\neq0$, propagation exists; however, due to the single root, it is anisotropic. Despite this directionality, linear response theory enables us to define the wave's group velocity as $\hat{c}_s = \nabla_k(\gamma\vec{C}_1\cdot\vec{k})$. That is, $\hat{c}_s\propto\gamma$, which is consistent with the behavior observed in Fig.~\ref{fig:fig03}(a).

From Eq.~\eqref{eq:disres}, one can infer that the aggregation coupling strength ($\alpha$) does not directly enable the system's propagating nature. Its influence, however, extends beyond this specific feature. In the limit of large spatial scales ($\vec{k}\rightarrow0$), the attenuation characterized by $\mathrm{Im}[\omega(\vec{k})]$ approaches $-\alpha$ (for $\alpha>0$). Then, if $\alpha = 0$, the \textit{damping} vanishes and thus a Goldstone-like mode emerges. That is, in essence, the aggregation coupling strength ($\alpha$) represents a \textit{damping gap}.

Exploring the system beyond the deterministic scenario ($\hat{G}_0^{-1} \hat{\mu}_0= \hat{\nu}$) allows us to characterize its observable response. Specifically, we derive an expression for the correlation function in the frequency domain via $\hat{C}(\vec{k}, \omega) = \langle\vec{\mu}(\vec{k}, \omega)\vec{\mu}(\vec{k}, \omega)^*\rangle$. Thus, since $\hat{G}_0$ is deterministic [Eq.~\eqref{eq:prop}], and the noise is white in nature, we obtain

\begin{equation}
    \hat{C}(\vec{k}, \omega) = \frac{2D_\nu}{(\alpha - \gamma\vec{k}\cdot\mathbf{C}_2\vec{k})^2 + (\gamma\vec{C}_1\cdot{\vec{k}}-\omega)^2},
    \label{eq:loren}
\end{equation}
which describes a Lorentzian function centered at $\omega=\gamma\vec{C}_1\cdot\vec{k}$, with width $(\alpha - \gamma\vec{k}\cdot\mathbf{C}_2\vec{k})$. 

Eq.~\eqref{eq:loren} suggests that, even during straight flight (with $D_\nu\neq0$), estimating the correlation function should reveal the system's propagating nature. However, to fully connect with observations in natural flocks~\cite{cavagna2025spin}, we need to discuss the effects of averaging $\hat{C}(\vec{k}, \omega)$ over all directions of $\vec{k}$. This procedure is defined as an integral: different Lorentzian functions are superimposed. Then, as terms with both $\vec{k}$ and $-\vec{k}$ are considered, the resulting $\langle\hat{C}(\vec{k}, \omega)\rangle$ is symmetric around $\omega = 0$. That is, although its profile might show signatures of wave propagation, indicators on the directional preference disappear. Remarkably, when the integral defining $\langle\hat{C}(\vec{k}, \omega)\rangle$ is evaluated under reasonable considerations ($\vec{k}\to0$), the approximate result is a single Lorentzian peak. This, in turn, erroneously suggests the system's incapability to sustain propagating waves.

\subsection{Spatial heterogeneity}
Note that even if we consider translationally invariant interaction rules [i.e., $\mathcal{A}(\vec{r}, \vec{r}\,') = \mathcal{A}(\vec{r}\,'-\vec{r})$] and rewrite $\vec{C}_1$ and $\mathbf{C}_2$ [Eq.~\eqref{eq:lap}] to remove the dependence on $(\vec{r}\,'-\vec{r})$, there remains an explicit dependence on $\vec{r}$ through $\rho$. To achieve spatial homogeneity, therefore, we not only need to assume that space is periodic but also that $\rho(\vec{r}) = \rho_0$ is constant. This latter assumption is reasonable when we explore the system's properties at a macroscopic scale ($\vec{k}\to 0$). However, beyond this limit, the effects of the dependence on $\vec{r}$ become noticeable. This regime is particularly relevant, as finite-size systems such as natural flocks fall within its scope.

The explicit space-dependent form of $\mathcal{L}$ complicates the extraction of clear conclusions. In light of this, we characterize the system's spatial heterogeneity as an effective small perturbation over the homogeneous scenario. The full Laplacian operator is thus defined as 
\begin{equation}
\mathcal{L}(\vec{r}) \approx \mathcal{L}_0 + \sigma \mathcal{L}_1(\vec{r}),
\label{eq:lap_pert}
\end{equation}
where $\sigma$ is a small constant factor. 

Within the perturbative scenario, the solution to Eq.~\eqref{eq:fdyn} is given by $\vec{\mu} \approx \vec{\mu}_0 + \sigma \vec{\mu}_1$. The zeroth-order approximation corresponds to the spatially-homogeneous system [Eq.~\eqref{eq:fdyn0}], which led to the derivation of Eqs.~\eqref{eq:disres} and~\eqref{eq:loren}. Dealing with the first order correction, on the other hand, allows us to find that $\vec{\mu}_1$ is characterized by $\hat{G}_0^{-1}\hat{\mu}_1 + \gamma\hat{\mathcal{L}}_1*\hat{\mu}_0 = 0$ (see Appendix~\ref{app:pert} for details). Thus, we obtain
\begin{equation}
    \hat{\mu} \approx \hat{\mu}_0 - \gamma\hat{G}_0(\hat{\mathcal{L}}_1* \hat{\mu}_0),
    \label{eq:born}
\end{equation}
which corresponds to a Born approximation with propagator $\hat{G}_0$. This solution has profound significance in relation to the system’s response. The term of spatial convolution, $(\hat{\mathcal{L}}_1* \hat{\mu}_0)$, implies that modes are not strictly decoupled. Therefore, evaluating the spectrum for velocity fluctuations will reveal interference rather than the system's preferred modes.

The effects described above can also be explained by focusing on the nature of interactions: non-reciprocity and heterogeneity generate non-normality. As the Laplacian in Eq.~\eqref{eq:fdyn} does not satisfy the commutation relation $\mathcal{L}\mathcal{L}^\dagger=\mathcal{L}^\dagger\mathcal{L}$, left and right eigenfunctions become distinct. Orthogonality breaks down; therefore, individual modes are no longer excited independently, and diagonalization in a Fourier basis becomes impossible. This is ultimately consistent with Eq.~\eqref{eq:born}.

\section{Measurement of correlations}\label{sec:corr}
We evaluate the spectral nature of our system using data obtained from simulations of Eqs.~\eqref{eq:iner}. Particularly, for our observations to be consistent with natural flocks, we explore a system with a comparable population size. 

Our analysis focuses on the spatiotemporal correlations among velocity fluctuations. The correlation function we use is defined as~\cite{cavagna2025spin}
\begin{equation}
    C(\vec{k}, t) = \frac{1}{N}\left\langle\sum_{i, j}^N \vec{\varphi}_i(t_0)\cdot\vec{\varphi}_j(t_0 + t)e^{-i\vec{k}\cdot\vec{r}_{ij}}\right\rangle_{t_0},
    \label{eq:corr}
\end{equation}
where $\vec{r}_{ij} = \vec{r}_i(t_0) - \vec{r}_j(t_0+t)$, with positions measured relative to the flock's instantaneous center of mass. The system's anisotropy is captured by the dependence of $C(\vec{k}, t)$ on the wave vector. Hence, we consider the significant spatial frequencies to be bounded by $k_0 = 2\pi/L$ and $k_{max} = \pi/\ell$, where $L$ is the size of the cubic simulation box and $\ell$ is the mean interparticle distance. To avoid ambiguities, we evaluate for the same effective wave numbers in all directional cases (as well as in the isotropic ones evaluated later). Finally, to assess the spectral features, we apply the Fourier transform in the time domain.

\begin{figure}
    \centering    \includegraphics[width=\linewidth]{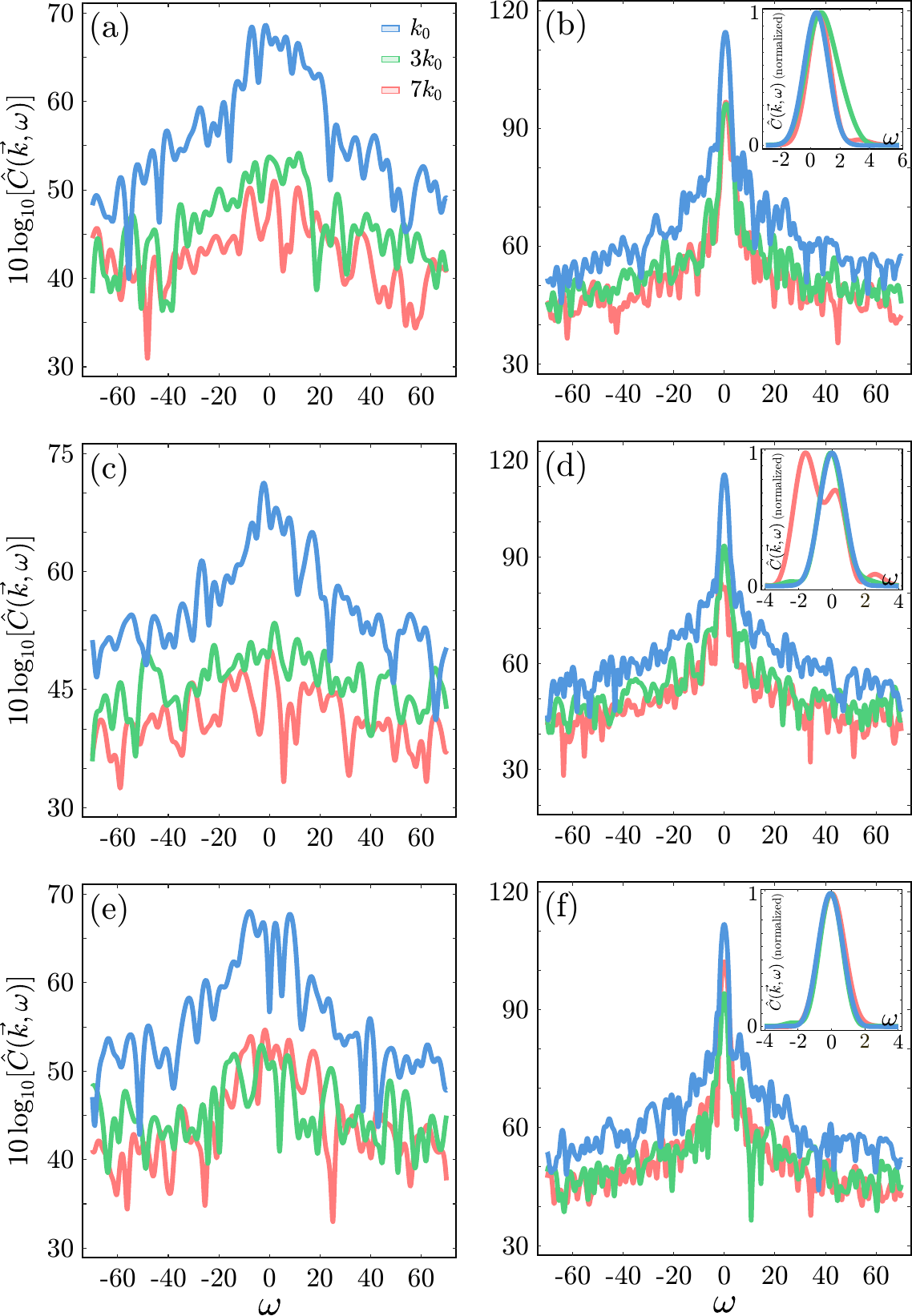}
    \caption{Spectra of spatiotemporal correlations for a simulation with parameters $(N, \alpha,  \gamma, D_\nu) = (500, 0.01, 50, 0.5)$. The probe wave vectors were defined as $\vec{k} = (k_x, 0, 0)$ for (a) and (b); $\vec{k} = (0, k_y, 0)$ for (c) and (d); and $\vec{k} = (0, 0, k_z)$ for (e) and (f). Left panels correspond to calculations using fluctuations emerging during the collective's straight flight; right panels correspond to calculations using fluctuations present during the turning event. The respective insets show magnifications of the normalized correlation functions. The significant wave numbers range between $k_0 = 0.063$ and $k_{max}=0.41$; we therefore considered a maximum scale factor of $7$. }
    \label{fig:fig04}
\end{figure}

Fig.~\ref{fig:fig04} reveals that signatures of our theoretical expectations [Eq.~\eqref{eq:loren}] are visible despite oscillations arising from using Eq.~\eqref{eq:corr}. For velocity fluctuations that emerge during the collective's straight flight [Figs.~\ref{fig:fig04}(a), \ref{fig:fig04}(c), and \ref{fig:fig04}(e)], the correlation functions display roughly a single maximum. This trend is particularly noticeable at the largest spatial scale ($|\vec{k}|=k_0$), aligning with the condition of homogeneity ($\vec{k}\to0$) that guarantees the validity of Eq.~\eqref{eq:loren}. Beyond $k_0$, and although similar profiles of $\hat{C}(\vec{k}, \omega)$ are maintained, a relation meeting $\omega=\gamma\vec{C}_1\cdot\vec{k}$ [Eq.~\eqref{eq:loren}] cannot be observed clearly. This behavior may be attributed to the increasing relevance of heterogeneity, as discussed for Eq.~\eqref{eq:born}. Note, additionally, that non-reciprocity gives rise to clusters of high and low activity that, however, are averaged out as Eq.~\eqref{eq:corr} does not account for such local variations.

\begin{figure}
    \centering
    \includegraphics[width=\linewidth]{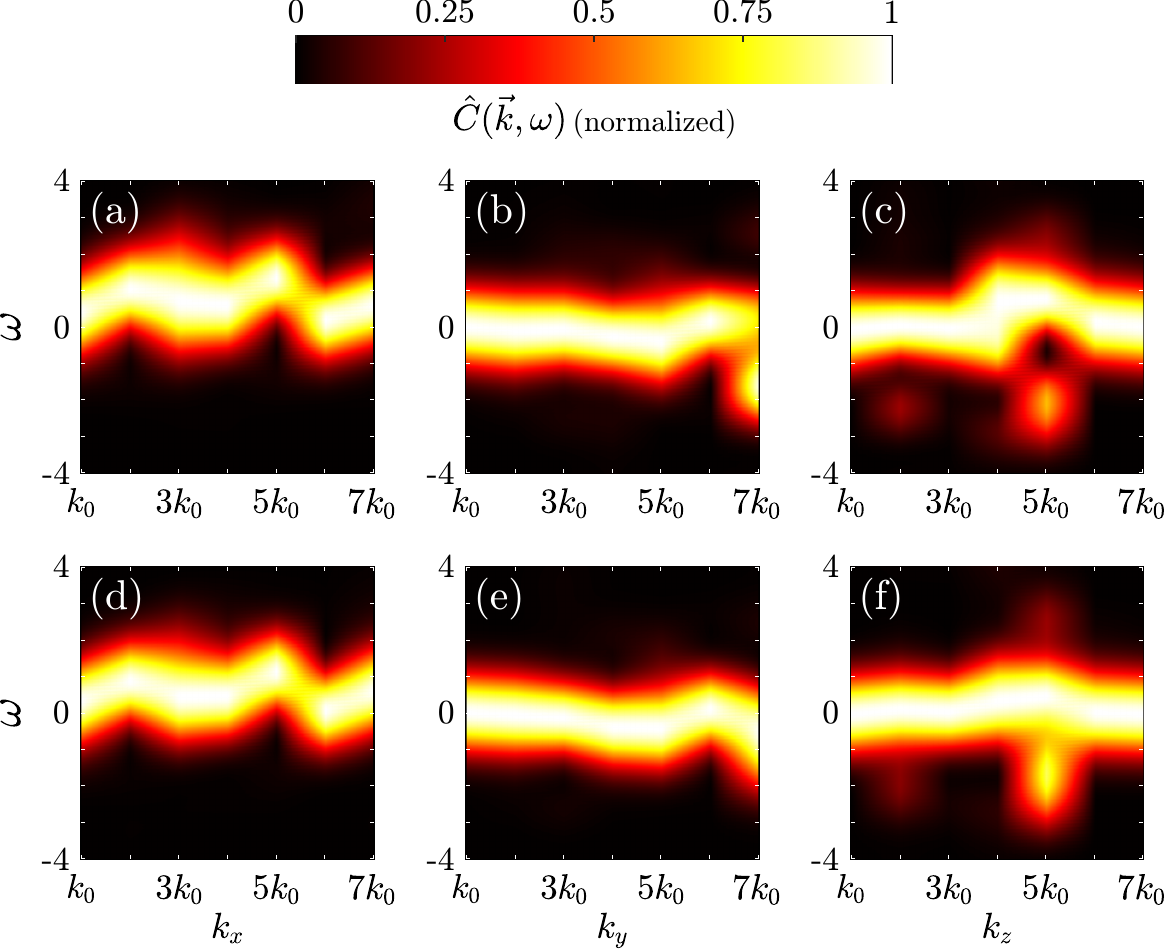}
    \caption{Dispersion diagrams calculated for two systems with parameters $(N, \alpha, \gamma, D_\nu) = (500, 0.01, 50, 0.5)$ (top row) and $(N, \alpha, \gamma, D_\nu) = (500, 0.01, 70, 0.5)$ (bottom row). The evaluated fluctuations were those present during the collective's turn; both systems started the event with identical configurations. The probe wave vectors were defined as $\vec{k} = (k_x, 0, 0)$ for (a) and (d); $\vec{k} = (0, k_y, 0)$ for (b) and (e); and $\vec{k} = (0, 0, k_z)$ for (c) and (f). The minimum wave number is $k_0=0.063$ and the spectra for intermediate values of $k$ are interpolations.}
    \label{fig:fig05}
\end{figure}

Focusing on the system's turning event [Figs.~\ref{fig:fig04}(b), \ref{fig:fig04}(d), and \ref{fig:fig04}(f)] yields quasi-Lorentzian profiles with well-defined peaks (see insets). However, these may simply emerge due to the initiator's sustained stimulus. Unlike white noise, the turning signal excites the system at specific frequencies. Fig.~\ref{fig:fig05}, which shows dispersion diagrams calculated using Eq.~\eqref{eq:corr}, is particularly revealing in this regard. In all panels, the brightest regions represent the system's preferred modes ($\vec{k}, \omega$). As is evident, then, although some hints of linearity are apparent near the maximum spatial scales ($k_0$), this behavior soon breaks down into intricate structures. That is, in spite of the coherent spectral profiles observed at specific wave numbers (Fig.~\ref{fig:fig04}'s insets), their evolution does not meet Eq.~\eqref{eq:loren}. Moreover, these dispersion diagrams do not capture the effects of varying $\gamma$ and, accordingly, they fail to provide an estimate of the wave's group velocity [Eq.~\eqref{eq:disres}].

\subsection{Isotropic averaging}
\begin{figure}
    \centering
    \includegraphics[width=\linewidth]{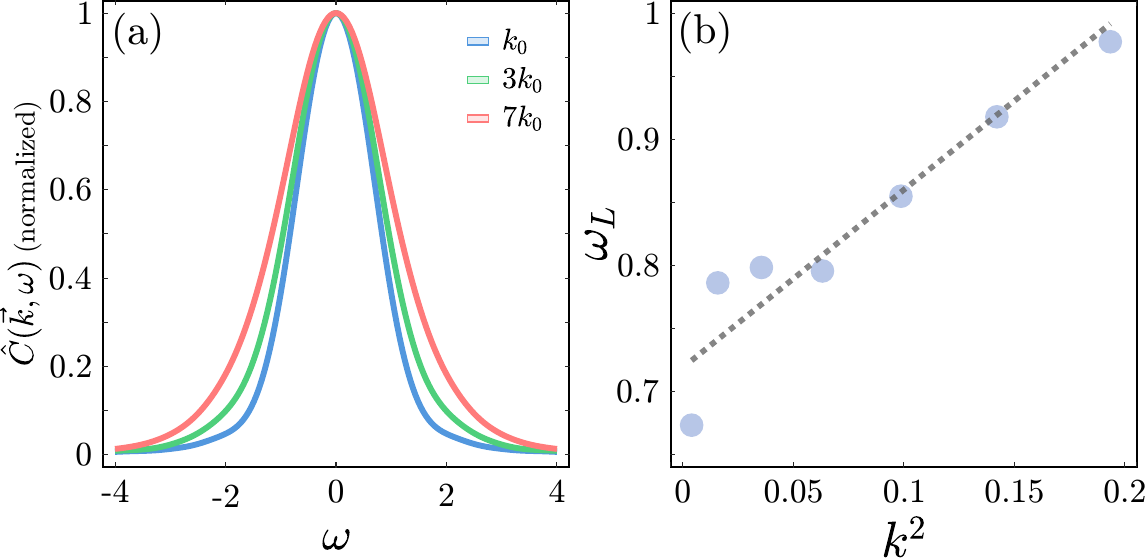}
    \caption{Spectral features obtained via isotropic averaging for a system with $(N, \alpha, \gamma, D_\nu) = (500, 0.01, 50, 0.5)$. Correlation functions in (a) were estimated for velocity fluctuations emerging during the collective's straight flight. In (b), respective half-width frequencies ($\omega_L$) are shown to scale approximately linearly with $k^2$. The dashed gray line represents a linear fit.}
    \label{fig:fig06}
\end{figure}

On top of the anomalies noted above, smooth quasi-Lorentzian profiles can be recovered if the explicit directional dependence in Eq.~\eqref{eq:corr} is relaxed ($\vec{k}\to k$). Fig.~\ref{fig:fig06} shows the results of calculating the isotropically averaged correlation function via
\begin{equation*}
    C(k, t) = \frac{1}{N}\left\langle\sum_{i, j}^N \vec{\varphi}_i(t_0)\cdot\vec{\varphi}_j(t_0 + t)\text{sinc}(kr_{ij})\right\rangle_{t_0},
\end{equation*}
where $r_{ij} = |\vec{r}_{ij}|$ and sinc(.) is the cardinal sine. 

The resulting spectra, displayed in Fig.~\ref{fig:fig06}(a), are consistent with the expectations discussed for $\langle\hat{C}(\vec{k}, \omega)\rangle$ [based on Eq.~\eqref{eq:loren}]. The isotropic average removes the directional features of $\hat{C}(\vec{k}, \omega)$, and, surprisingly, it also eliminates the effects of heterogeneity. Remarkably, as shown in Fig.~\ref{fig:fig06}(b), the resulting profiles show scaling features characteristic of overdamped systems.

\section{Discussion}\label{sec:dis}
The model we presented in this paper reproduces information-transfer features exhibited by natural flocks. From a microscopic perspective, we demonstrated that the turning signal generated by a single individual is transmitted with minimal loss throughout the entire collective (Fig.~\ref{fig:fig02}). In particular, we found that the turning information propagates at a constant speed, in a ballistic-like fashion (Fig.~\ref{fig:fig03}). While turning, individuals experience only minor speed fluctuations and follow equal-radii trajectories (Fig.~\ref{fig:fig01}). Remarkably, all these features, characteristic of starling murmurations~\cite{attanasi2014information, attanasi2015emergence, cavagna2025spin}, emerge despite our system's lack of explicit inertial components.

To understand the mechanism driving the propagation of information, we analyzed our model in the continuum limit [Eq.~\eqref{eq:fdyn}]. This then revealed that the system's propagating features are enabled by the non-reciprocity of interactions. Remarkably, despite this capability, the system cannot be strictly categorized as underdamped. The successful completion of a collective turn also depends on consensus modes, characterized by purely real eigenvalues. Ultimately, the initiator does not perturb the system with a single impulse, but applies a sustained stimulus.

Within the continuum context, further investigation led us to recognize the emergence of spectral anomalies. These, rooted in the non-normal nature of $\mathcal{L}$ [Eq.~\eqref{eq:fdyn}]. To demonstrate the emergence of these effects, we considered spatial heterogeneity as an effective correction to the homogeneous scenario. As a result, a phenomenon of mode coupling was evidenced through a component of spatial convolution [Eq.~\eqref{eq:born}]. Notably, as the approximate solution for $\vec{\mu}$ follows a Born approximation, the resulting effects could be interpreted as scattering. As a side note, the time dependence of the interaction network could also be included as a small perturbation. However, we believe that such a treatment would have obscured our conclusions. 

It is worth remarking that achieving spatial homogeneity in finite-size collectives, considering moreover non-trivial interaction rules, is challenging. Heterogeneity, in contrast, emerges naturally as a product of individual states, non-periodic boundary conditions, or both. Consequently, even if the system self-organizes into a perfect crystal with uniform properties, the lack of periodic boundaries alone would induce mode coupling. This distinction is crucial, as natural flocks are essentially systems with finite, open boundaries where individual roles are also position-dependent~\cite{attanasi2015emergence}. 

To maintain coherence with natural flocks, our estimates of the spatiotemporal correlation functions were performed within the finite-size regime. Even though the resulting spectra [Fig.~\ref{fig:fig04}] exhibited signs of consistency with our theoretical expectations, anomalies inherent to non-ideal conditions were also noticeable. To remove undesirable effects, we then averaged the spectra isotropically. Strikingly, the results displayed overdamped-like profiles (Fig.~\ref{fig:fig06}), masking signatures of propagation and heterogeneity. This outcome is particularly relevant as observations in natural flocks have reached a similar conclusion~\cite{cavagna2025spin}.

While our primary objective was to analyze the nature of collective turns, our model offers several opportunities for future work. For instance, we have not yet explored phase transitions or their dependence on the model parameters. Spectral estimations could also be improved by simulating considerably larger systems and highly idealized conditions, analogous to the approach in~\cite{tu1998sound}. Although our simulations with fixed $\eta$ revealed the underlying nature of our system, more realistic interaction rules may capture specific dynamics not considered here. Further considerations may also be important for the sake of realism. For example, despite its successful application in characterizing other systems~\cite{attanasi2014finite, de2025stochastic}, the mean-field nature of the aggregation mechanism (also discussed in~\cite{lizarraga2025swarming}) might not be the optimal formulation. Similarly, different implementations of the turning protocol could give rise to effects obscured in our simulations. Beyond these specific considerations, given the flexibility of its formulation, we believe our model could be tailored to characterize other natural collectives, such as fish or insects~\cite{mugica2022scale, de2025stochastic, sayin2025behavioral, szabo2006phase}, in a physically consistent manner.

The \textit{damping paradox} addressed in~\cite{cavagna2025spin} arises from discrepancies between the behavior of real flocks and the ISM nature. Accordingly, the authors showed that adding a quartic term to the ISM alignment dynamics may solve the inconsistencies. While this modification is theoretically sound, it increases the model's complexity (and, consequently, its functional interpretation). Our system, in a complementary manner, shows that a fundamental behavioral feature, within an alternative formulation, could be sufficient to untangle the apparent contradictions. Certainly, further empirical validation is needed to determine whether our model can quantitatively capture the behavior of natural flocks. Nevertheless, even as a toy model, it provides a different perspective on the flocking phenomenon. As a final remark, it is worth noting that non-reciprocity has been shown to play a significant role in the behavior of diverse active matter systems~\cite{you2020nonreciprocity, fruchart2021non, bowick2022symmetry, dinelli2023non, tan2022odd, rouzaire2026dynamics, hang2026self}. Therefore, additional research may elucidate the connection between our system and concepts such as transient growth, exceptional points, or the non-Hermitian skin effect. 

\begin{acknowledgments}

M.A.M.A. would like to thank FAPESP, grant 2021/14335-0, and CNPq, grant 303814/2023-3 for financial support.

\end{acknowledgments}
\appendix

\section{Numerics}
\label{app:num}
In all simulations, individuals were initially uniformly distributed in a cubic box with a side length of $100$. Initial velocities were drawn from uniform distributions on the intervals $[0, 13]$ ($x$-axis), $[0, 10]$ ($y$-axis), and $[-12, 12]$ ($z$-axis), and $\eta$ was fixed at 6. Apart from the simulations associated with Fig.~\ref{fig:fig01}, all others were performed over $35\times10^3$ time steps. The system was initially driven solely by noise during $12.5\times10^3$ time steps, followed by $\pi/2$ turning events. These turns lasted for $5\times10^3$ time steps, after which the individuals resumed straight flight.

The model was numerically integrated using the Euler-Maruyama method with a time step of $\Delta t=10^{-3}$. Time ($t$), in all figures, was reported in simulation units. Larger values of $\Delta t$ gave rise to a \textit{ringing} phenomenon, and consequently, led to the appearance of two artificial peaks in the frequency domain. 

Periodic boundary conditions (PBCs) were applied in simulations to estimate correlation functions. The size of the box ($L = 100$) was chosen mainly to avoid a disruption in information flow. Conversely, open boundary conditions (OBCs) were used to calculate propagation speeds based on individual orientations. The evaluation of unfolded trajectories from simulations with PBCs also revealed linear regimes similar to those shown in Fig.~\ref{fig:fig03}(a). Likewise, correlation functions obtained from simulations with OBCs or PBCs (with the box size considerably larger than the collective's linear size), resulted in profiles similar to those shown in the main text. 

For calculations regarding the estimation of propagation speeds based on individual orientations, velocities were smoothed using a second-order Butterworth filter. This procedure was performed mainly to remove fluctuations inherent to the discrete computation. No filters were used in any frequency-related figures; we addressed spectral leakage through windowing.

\begin{figure}
    \centering
    \includegraphics[width=\linewidth]{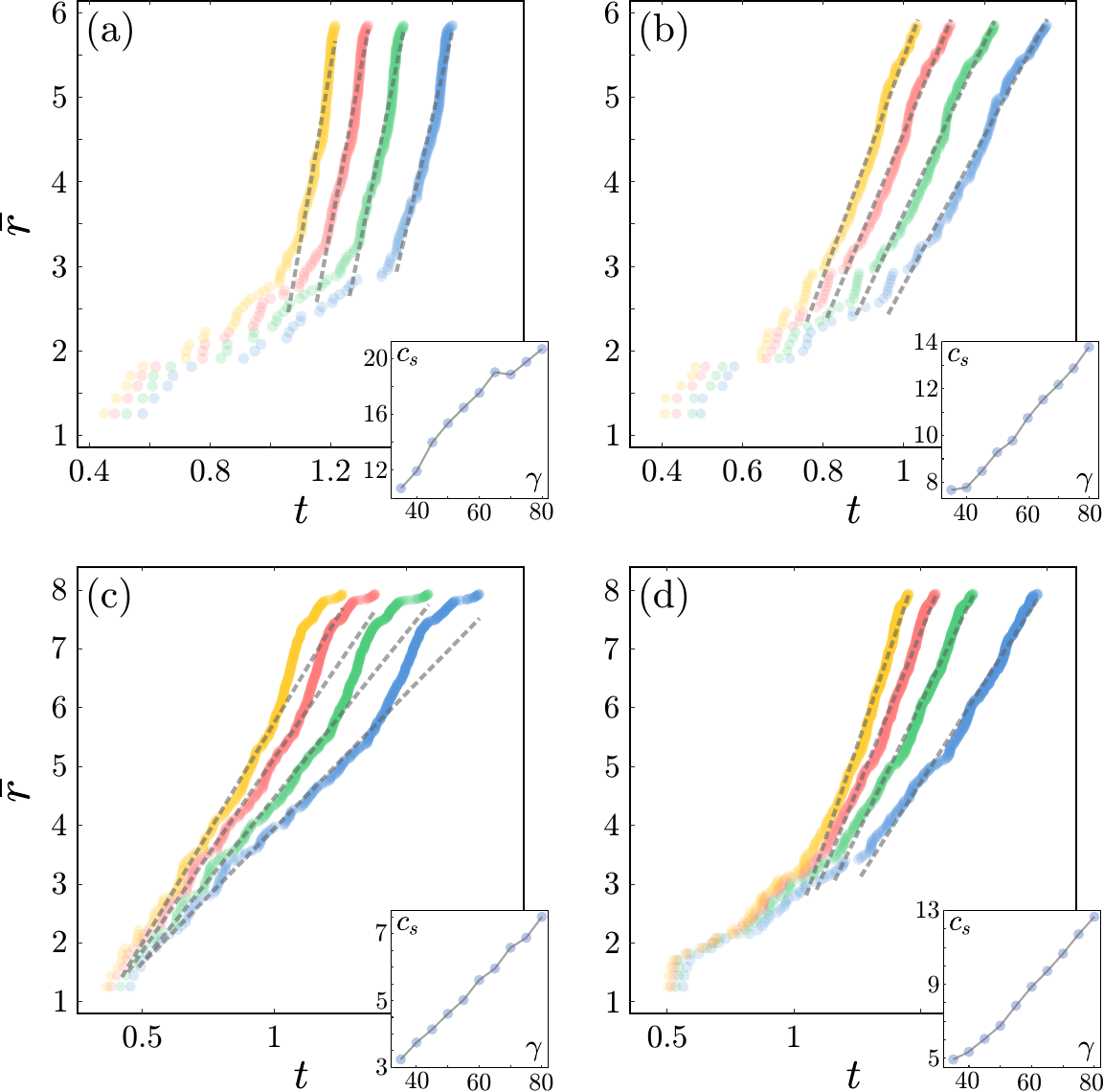}
    \caption{Propagation speeds estimated for systems of $N = 200$ (top row) and $N = 500$ (bottom row) individuals. The aggregation coupling strength was set to $\alpha = 0.01$ (left column) and $\alpha = 1$ (right column). In each panel, turning events started in systems with the same configuration; colored lines represent results for $\gamma = 50$ (blue), $\gamma = 60$ (green), $\gamma = 70$ (red), and $\gamma = 80$ (yellow). In contrast to Fig.~\ref{fig:fig03}(a), here only the data point corresponding to the initiator was excluded. The insets show the relation between $\gamma$ and estimated $c_s$ considering additional cases.}
    \label{fig:figAPP}
\end{figure}

Additional results, complementing Fig.~\ref{fig:fig03}, are shown in Fig.~\ref{fig:figAPP}.

\section{The continuum approximation}
\label{app:cont}
Approximating the system's dynamics towards the continuum limit follows a standard procedure. We start by rewriting Eq.~\eqref{eq:iner} as
\begin{equation*}
    \dot{\vec{\varphi}}_i = \vec{\nu}_i- \alpha \vec{\varphi}_i + \gamma\sum_jA_{ij}(\vec{\varphi}_j - \vec{\varphi}_i) - \frac{\gamma}{N}\sum_k\sum_j A_{kj}(\vec{\varphi}_j - \vec{\varphi}_k),
\end{equation*}
such that, by defining the relations
\begin{align}
        \sum_jA_{ij}(\vec{\varphi}_j - \vec{\varphi}_i) &= -\sum_j L_{ij}\vec{\varphi}_j,\label{eq:SM01}\\
        \sum_k\sum_j A_{kj}(\vec{\varphi}_j - \vec{\varphi}_k) &= \sum_j\vec{\varphi}_j (d_j^{out} -d_{j}^{in}),\label{eq:SM02}    
\end{align}
we arrive at Eq.~\eqref{eq:fluc}.

\subsection{Field of interactions}
Eq.~\eqref{eq:SM01} gives rise to the Laplacian operator $\mathcal{L}$. In the continuum limit, we assume that distances between neighboring individuals, 
\begin{equation*}
\vec{r}_{ji} = \vec{r}_j - \vec{r}_i,  
\end{equation*}
 are small. Then, under such condition, we define the field of velocity fluctuations $\vec{\mu}(\vec{r}, t)$. 

At a specific instant in time, we take advantage of the system's discrete structure to establish the relation $\vec{\mu}(\vec{r}_j) = \vec{\mu}(\vec{r}_i + \vec{r}_{ji})$. Thus, given that $\vec{r}_{ji} \rightarrow0$, performing a Taylor expansion to the second order leads to
\begin{equation*}
    \vec{\mu}(\vec{r}_i + \vec{r}_{ji}) \approx \vec{\mu}(\vec{r}_i) + (\vec{r}_{ji}\nabla)\vec{\mu}(\vec{r}_i) + \frac{1}{2}(\vec{r}_{ji}\nabla)^2\vec{\mu}(\vec{r}_i).
\end{equation*}
That is, the difference between velocity fluctuations at specific positions in space can be characterized by 
\begin{equation}
    \vec{\mu}(\vec{r}_j) - \vec{\mu}(\vec{r}_i) \approx  (\vec{r}_{ji}\nabla)\vec{\mu}(\vec{r}_i) + \frac{1}{2}(\vec{r}_{ji}\nabla)^2\vec{\mu}(\vec{r}_i).
    \label{eq:SM03}
\end{equation}
Then, by plugging Eq.~\eqref{eq:SM03} into Eq.~\eqref{eq:SM01}, we find that the interaction term follows
\begin{equation*}
     \sum_j A_{ij}[\vec{\mu}(\vec{r}_j) - \vec{\mu}(\vec{r}_i)]= \Gamma_{1, i} + \Gamma_{2, i},
\end{equation*}
where
\begin{align}
    \Gamma_{1, i} &= \sum_jA_{ij}(\vec{r}_{ji}\nabla)\vec{\mu}(\vec{r}_i), \label{eq:SM04}\\
    \Gamma_{2, i} &= \frac{1}{2}\sum_jA_{ij}(\vec{r}_{ji}\nabla)^2\vec{\mu}(\vec{r}_i). \label{eq:SM05}
\end{align}

Notice, from Eq.~\eqref{eq:SM04}, that we can define an additional variable 
\begin{equation}
    \zeta_{1, i} = \sum_jA_{ij}\vec{r}_{ji},
    \label{eq:zeta1}
\end{equation}
such that $\Gamma_{1,i} = (\zeta_{1, i}\nabla)\vec{\mu}(\vec{r}_i)$. In the continuum, then, performing a procedure of coarse-graining over $\zeta_{1, i}$ leads to $-\vec{C}_1$ of Eq.~\eqref{eq:lap} [the negative sign appears simply to ensure consistency with Eq.~\eqref{eq:SM02}]. Finally, by removing the index, which is characteristic of the system's discrete structure, Eq.~\eqref{eq:SM04} becomes $\bar{\Gamma}_1 =- (\vec{C}_1\nabla)\vec{\mu}(\vec{r})$.

To deal with Eq.~\eqref{eq:SM05}, we use a procedure similar to that employed for Eq.~\eqref{eq:SM04}. The main difference, in this case, is that the quadratic term $(\vec{r}_{ji}\nabla)^2$ gives rise to 
\begin{equation*}
    \zeta_{2, i} = \frac{1}{2}\sum_jA_{ij}(\vec{r}_{ji}\otimes \vec{r}_{ji}).
\end{equation*}
Under coarse-graining, we thus obtain $-\mathbf{C}_2$ introduced in Eq.~\eqref{eq:lap} (the factor $1/2$ was absorbed for notational simplicity). Consequently, Eq.~\eqref{eq:SM05} yields $\bar{\Gamma}_2 = -\nabla\cdot(\mathbf{C}_2\nabla)\vec{\mu}(\vec{r})$. 

\subsection{Global stimulus}
Eq.~\eqref{eq:SM02} corresponds to $\vec{\xi}_i$ in Eq.~\eqref{eq:fluc}. In the continuum limit, it becomes
\begin{equation*}
    \vec{\xi}({\vec{r}, t}) = \frac{\gamma}{N}\int d\vec{r}\,' \vec{\mu}(\vec{r}\,', t) D(\vec{r}\,'),
\end{equation*}
where $D(\vec{r}\,')$ characterizes the local asymmetry of interactions. Since the integrand does not depend on $\vec{r}$ but only on $t$, the spatially uniform field can be written as: $\vec{\xi}(\vec{r}, t) = \vec{\xi}(t)$. Consequently, applying the spatial Fourier transform yields
\begin{equation*}
    \hat{\xi}(\vec{k}, t) = \vec{\xi}(t)\int  e^{-i\vec{k}\cdot\vec{r}}d\vec{r},
\end{equation*}
which is the same as $\hat{\xi}(\vec{k}, t)\propto\vec{\xi}(t)\delta(\vec{k})$. Therefore, in the regime considered for our analysis ($\vec{k}\neq 0$), the term $\hat{\xi}(\vec{k}, t)$ becomes zero.

\section{Polarization and propagation}\label{app:pp}
Consider the polarization order parameter $\bar{v}_a = 1/N\left|\sum_i\dot{\vec{r}}_i\right|$, where the normalization $(\dot{\vec{r}}_i/|\dot{\vec{r}}_i|)$ was neglected for simplicity. Then, by plugging Eq.~\eqref{eq:agg} into $\bar{v}_a$, with $\alpha\to0$, we obtain:
\begin{equation*}
    \bar{v}_a \approx \frac{1}{N}\left|\sum_i\vec{f}_i\right|.
\end{equation*}
This expression can be rewritten using the integral of $\dot{\vec{f}}_i$ [Eq.~\eqref{eq:ali}] as
\begin{equation*}
    \bar{v}_a \approx \left|\int \frac{1}{N}\sum_i\vec{\nu}_i dt + \frac{\gamma}{N}\sum_i\sum_jA_{ij}(\vec{r}_j - \vec{r}_i) + \vec{C}\right|,
\end{equation*}
where $\vec{C}$ is a constant of integration. Finally, by employing Eq.~\eqref{eq:zeta1}, we arrive at
\begin{equation}
    \bar{v}_a \approx \left|\int \frac{1}{N}\sum_i\vec{\nu}_i dt + \frac{\gamma}{N}\sum_i\zeta_{1,i} + \vec{C}\right|.
    \label{eq:pol}
\end{equation}

It is worth noting that the term involving $\vec{\nu}_i$ in Eq.~\eqref{eq:pol} vanishes in the thermodynamic limit. More importantly, although the average $1/N \sum_i\zeta_{1,i}$ does not exactly match the definition of $\vec{C}_1$ [Eq.~\eqref{eq:lap}], it constitutes an effective analog. Therefore, as the propagation speed ($\hat{c}_s$) is characterized by $\gamma\vec{C}_1$ [Eq.~\eqref{eq:disres}], we can argue from Eq.~\eqref{eq:pol} that $\bar{v}_a\propto\hat{c}_s$. This relationship is demonstrated in Fig.~\ref{fig:fig03}(b), where $c_s$ and $(1-v_a)^{-1/2}$ were explicitly chosen as axes to allow a direct comparison with results reported for natural flocks~\cite{attanasi2014information}. 

\section{Perturbative treatment}\label{app:pert}
In Eq.~\eqref{eq:lap_pert}, we approximate the Laplacian operator ($\mathcal{L}$) as the sum of a homogeneous component ($\mathcal{L}_0$), and a heterogeneous one ($\sigma\mathcal{L}_1$). In turn, the system's solution is approximated as $\vec{\mu} \approx \vec{\mu}_0 + \sigma\vec{\mu}_1$. In the zeroth-order, we have
\begin{equation*}
    \dot{\vec{\mu}}_0 + \alpha\vec{\mu}_0 + \gamma\mathcal{L}_0\vec{\mu}_0 + \vec{\xi} = \vec{\nu},
\end{equation*}
which explicitly describes the system under spatial homogeneity. Then, by transforming this equation to the frequency domain, we obtain  
\begin{equation*}
    \left[\alpha - \gamma\vec{k}\cdot \mathbf{C}_2\vec{k}+ i(\gamma\vec{C}_1\cdot \vec{k} - \omega)\right]\vec{\mu}_0 = \vec{\nu}, 
\end{equation*}
where we have already removed $\vec{\xi}$ as our interest lies in $\vec{k}\neq 0$. In the main text, this equation is written in terms of the propagator as $\hat{G}_0^{-1} \hat{\mu}_0= \hat{\nu}$.

In the first-order approximation, we plug the approximate $\mathcal{L}$ and $\vec{\mu}$ into the continuum dynamics [Eq.~\eqref{eq:fdyn}]. Thus, we obtain 
\begin{equation*}
(\dot{\vec{\mu}}_0 + \sigma\dot{\vec{\mu}}_1) + \alpha(\vec{\mu}_0 + \sigma\vec{\mu}_1) + \gamma(\mathcal{L}_0 + \sigma\mathcal{L}_1)(\vec{\mu}_0 + \sigma\vec{\mu}_1) = \vec{\nu}-\vec{\xi},
\end{equation*}
which, by grouping in powers of $\sigma$, leads to
\begin{equation*}
\dot{\vec{\mu}}_1 + \alpha\vec{\mu}_1 + \gamma\mathcal{L}_0\vec{\mu}_1 + \gamma\mathcal{L}_1\vec{\mu}_0 = 0.
\end{equation*}
Note that we have neglected the higher order term multiplying $\sigma^2$. Finally, in the frequency domain, this equation becomes
\begin{equation*}
    \left[\alpha - \gamma\vec{k}\cdot \mathbf{C}_2\vec{k}+ i(\gamma\vec{C}_1\cdot \vec{k} - \omega)\right]\hat{\mu}_1 + \gamma\hat{\mathcal{L}}_1*\hat{\mu}_0 = 0, 
\end{equation*}
which yields Eq.~\eqref{eq:born}.

\bibliography{apssamp}

\end{document}